\documentclass[nofootinbib,aps,showpacs,preprint]{revtex4}%
\usepackage{amsfonts}
\usepackage{amsmath}
\usepackage{amssymb}
\usepackage{graphicx}%
\begin{document}
\preprint{ }
\title{Electric Dipole Moment of Dirac Fermionic Dark Matter}
\author{Jae Ho Heo}
\email{jheo1@uic.edu}
\affiliation{Physics Department, University of Illinois at Chicago, Chicago, Illinois
60607, USA }

\begin{abstract}
The direct limit of electric dipole moment (EDM) and direct search for dark
matter by EDM interaction are considered as including the electromagnetic
nuclear form factor, in case that the dark matter candidate is a Dirac
particle. The WIMP electric dipole moment constrained by direct searches must
be lower than $7\times10^{-22}e$ $cm$ for WIMP mass of 100 GeV to satisfy the
current experimental exclusion limits at XENON10 and CDMS II. We also consider
the CP violation of EDM and the WIMP discovery by EDM intereaction in the future.

\end{abstract}

\maketitle

\section{INTRODUCTION}

Dark matter (DM) has been postulated to explain various observations from
gravitational effects on visible matter and plays an important role to explain
structure formation and galaxy evolution. In recent years observations and the
high precision analysis of the cosmic microwave background radiation have
provided spectacular confirmation of the astrophysical evidence for DM, that
is to say about quarter of the energy density of the universe is dark matter.
Dark matter appears to be consisting of nonrelativistic particles that only
interact gravitationally and perhaps by weak interaction. Mostly the coupling
to photons is assumed to be nonexistent or very weak, so the electromagnetic
interactions have not been considered seriously. A possible scenario for
electromagnetic interaction of Dirac fermionic dark matter with nonzero
magnetic dipole moment \cite{heo09} has been considered in the standard model
context. No additional particles are assumed except for the DM candidate near
electroweak scale $(10\sim1000$ GeV$)$. The various experimental bounds was
investigated \cite{mpos00,ksig04,spro07} in the past and a new experimental
technique \cite{sgad08} for DM detection by electromagnetic interaction has
been suggested for nonzero dipoles. There is a DM scenario for the
electromagnetic interaction by the fractional or millie charged particles
\cite{lbo82,sdv00}, but the DM scenario is for the case that another massless
$U(1)$ gauge boson, called paraphoton, exists beyond the standard model.

In this letter we consider electric dipole moment effect of Dirac fermionic
dark matter. The electric dipole moment (EDM) constrained by direct searches
is investigated as considering the electromagnetic form factor of the nucleus.
The WIMP-nucleus elastic scattering is due to spin independent interaction,
that gives the WIMP electric dipole moment very strict bound since the
WIMP-nucleus elastic scattering cross sections are enhanced by the square of
nuclear charge (number of protons in the nucleus), $Z^{2}$. The WIMP electric
dipole moment constrained by direct searches must be lower than $7\times
10^{-22}e$ $cm$ for WIMP mass of 100 GeV to satisfy the current experimental
exclusion limits. WIMP electric dipole moment is scaled as considering the
current experimental exclusion limit of electric dipole moment (EDM)
\cite{Bcr02} for the known Dirac particles and the scenario that presented in
Ref.\cite{heo09} to investigate the WIMP detectability. A simple model
(Lagrangian) with the complex dipole coupling is also introduced. Although we
consider that the interaction of EDM is suppressed by CP violation, the
suppression could be compensated by the enhancement of spin independent
interaction. WIMP could thus be detected by the EDM interaction in near future
if the suppression is not seriously small.

\section{Constraints from Direct Searchs}

The detection of dark matter is controlled by their elastic scattering with a
nucleus in a detector. In this case the $t$-channel exchange of a
photon\footnote{A $Z$-boson exchange is also possible in the DM scenario of
Ref.\cite{heo09}, but the interaction by a $Z$-boson exchange is negligible
for the low momentum transfer.} is only possible as shown in Fig. 1, since the
two interacting particles are distinguishable.%

%TCIMACRO{\FRAME{ftbpFU}{8.0814cm}{2.5481cm}{0pt}{\Qcb{Feynman diagram relevant
%to WIMP-nucleon elastic scattering. The hatched circle indicates the vertex
%for the dipole coupling.}}{}{fig1.eps}{\special{ language "Scientific Word";
%type "GRAPHIC";  maintain-aspect-ratio TRUE;  display "USEDEF";
%valid_file "F";  width 8.0814cm;  height 2.5481cm;  depth 0pt;
%original-width 5.5166in;  original-height 1.8922in;  cropleft "0";
%croptop "0.9027";  cropright "1";  cropbottom "0";
%filename 'FIG1.eps';file-properties "XNPEU";}}}%
%BeginExpansion
\begin{figure}
[ptb]
\begin{center}
\includegraphics[
trim=0.000000in 0.000000in 0.000000in 0.184111in,
height=2.5481cm,
width=8.0814cm
]%
{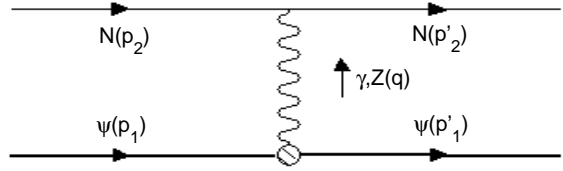}%
\caption{Feynman diagram relevant to WIMP-nucleon elastic scattering. The
hatched circle indicates the vertex for the dipole coupling.}%
\end{center}
\end{figure}
%EndExpansion

The effective Lagrangian at the quark level may be described by%

\begin{equation}
\mathcal{L}_{q}=\frac{dee_{q}}{\boldsymbol{q}^{2}}\left(  \overline{\psi
}\sigma^{\mu\nu}q_{\nu}\gamma^{5}\psi\right)  \left(  \overline{q}\gamma_{\mu
}q\right)  ,
\end{equation}
where $d$ is the WIMP electric dipole moment, $e$ is the electric coupling and
$e_{q}$ is an electric charge for the quark $q$.

In nonrelativistic case, the bi-spinor products may be expanded with respect
to the low momentum transfer. In the leading order of the momentum transfer,
only time component is taken.

The effective interaction of WIMP with a nucleon is%

\begin{equation}
\mathcal{L}_{N}\supset2\frac{def_{N}}{\boldsymbol{q}^{2}}M\left(
\overline{\psi}\gamma^{5}\psi\right)  \left(  N^{\dagger}N\right)  ,
\end{equation}
where $N$ stands for the proton (p) or neutron (n) and $f_{N}$ is the
effective coupling of WIMP to nucleons. In this case, $f_{p}=2e_{u}%
+e_{d}=1,f_{n}=e_{u}+2e_{d}=0$. The contributions of the heavy quark ($c,b$
and $t$) to WIMP-nucleon cross section can be related to the gluon
contribution by QCD, but those appear in quark-antiquark pairs. So three
valance quarks give the nucleon its global electric charge and the
contributions of each valance quark in the nucleon add coherently.

The similar argument can be applied to the nuclei. The constructive coherent
interactions give the squared scattering amplitude of the nucleus%

\begin{equation}
|\mathcal{M}|^{2}=16\frac{d^{2}(Ze)^{2}M^{2}m^{2}}{\boldsymbol{q}^{4}}\frac
{1}{2S+1}\sum\limits_{\text{spins}}(\boldsymbol{q\cdot S})^{2}F^{2}%
(|\boldsymbol{q}|)\text{,}%
\end{equation}
where $S=\frac{1}{2}$ is the WIMP spin, $Z$ is the nuclear electric charge
(the number of protons in the nucleus) and $F^{2}(|\boldsymbol{q}|)$ is the
electromagnetic form factor that is related to the electric charge
distribution in a nucleus. For the low momentum transfer, we cannot consider
the large nuclei as a point particle. We need consider the charge distribution
of the nucleons in a nucleus. We take the Helm form factor \cite{jeng92} that
was introduced as a modification of the form factor for an uniform sphere
multiplied by a gaussian to account for the soft edge of the nucleus.%

\begin{equation}
F^{2}(|\boldsymbol{q}|)=\left[  \frac{3j_{1}\left(  |\boldsymbol{q}%
|R_{1}\right)  }{|\boldsymbol{q}|R_{1}}\right]  ^{2}\exp\left[  -\left(
|\boldsymbol{q}|s\right)  ^{2}\right]  ,
\end{equation}
where%

\[
j_{1}(x)=\frac{\sin x}{x^{2}}-\frac{\cos x}{x}%
\]
is a spherical Bessel function of the first kind, and where $s\simeq1$ fm is
the nuclear skin thickness and $R_{1}=\sqrt{R^{2}-5s^{2}}$ is an effective
nuclear radius for nuclear radius $R\simeq1.2$ fm $A^{1/3}$ for an atomic
nucleus of atomic number $A$. This form factor is often referred as the
"Woods-Saxon" form factor though this form factor is not that obtained from
the Fourier transform of the Woods-Saxon density distribution.

The WIMP-nucleus differential cross section results in%

\begin{equation}
\ \ \frac{d\sigma_{el}}{d\boldsymbol{q}^{2}}=\frac{2\alpha Z^{2}d^{2}}%
{v^{2}\boldsymbol{q}^{2}}F^{2}(|\boldsymbol{q}|),
\end{equation}
where $\alpha=e^{2}/4\pi\simeq1/137$ is the electric fine structure constant.
The cross section has the Coulomb-like singularity at $\boldsymbol{q}^{2}=0$.

The expected event rate depends on WIMP-nucleus cross section and the WIMP
flux on the Earth, so the event rate per unit target mass and unit time is%

\begin{equation}
\frac{dR}{dE_{R}}=\frac{\rho_{D}}{mM}\int vf(v)\frac{d\sigma_{el}}{dE_{R}%
}dv\text{,}%
\end{equation}
where $\rho_{D}\simeq0.3$ GeV/cm$^{3}$ is the local DM density in the solar
vicinity and $f(v)$ is the WIMP velocity distribution funtion in the frame of
the detector.

With the relation $\boldsymbol{q}^{2}$ $=2mE_{R}$, the differential event rate is%

\begin{equation}
\frac{dR}{dE_{R}}=\frac{\alpha Z^{2}d^{2}\rho_{D}}{mM}\frac{F^{2}(E_{R}%
)}{E_{R}}\int\frac{f(v)}{v}dv\text{.}%
\end{equation}

The final formula for the event rate per unit mass and unit time is given by%

\begin{equation}
R=\frac{\alpha Z^{2}d^{2}\rho_{D}}{mM}\int_{E_{R,\min}}^{E_{R,\max}}%
dE_{R}\frac{F^{2}(E_{R})}{E_{R}}\cdot\frac{1}{2v_{E}}\left[
\operatorname{erf}\left(  \frac{v_{\min}+v_{E}}{v_{0}}\right)
-\operatorname{erf}\left(  \frac{v_{\min}-v_{E}}{v_{0}}\right)  \right]  ,
\end{equation}
with $v_{\min}=\sqrt{\frac{E_{R}m}{2M_{r}^{2}}}$, $v_{0}=220$ km/s is the
circular speed of the Sun around the Galactic center and $v_{E}=232$ km/s is
the average velocity considered the Earth speed to the Sun. \ Considering the
recoil energy ranges at Table I, the electric dipole moment of a WIMP
constrained by direct searches must be lower than $7\times10^{-22}e$ $cm$ for
WIMP mass of 100 GeV.

\begin{table}[t]
\caption{Current and planned Dark Matter detectors. XENON10 and CDMS II are
the current detectors and SuperCDMS is a planned detector.}%
\begin{ruledtabular}
\begin{tabular}{ccccc}
$\text{Experiment}$ & $\text{Recoil energy range}$  & $\text{Target}$& $\text{Nuclear charge (Z)}$ & $\text{
Mass}$ \\
\hline
$\text{XENON10}$ & $4.5\sim 27 \text{KeV}$ & $^{131}\text{Xe}$ & $54$ & $5.4 \text {Kg}$ \\
$\text{CDMS II}$ & $10\sim 100 \text{KeV}$ & $^{73}\text{Ge}$ & $32$ & $100\text {Kg}$
\\
$\text{SuperCDMS}$ & $15\sim 45 \text{KeV}$ & $^{73}\text{Ge}$&$32$ & $100 \text{
Kg}$\\
\end{tabular}
\end{ruledtabular}\end{table}

\section{Electric Dipole effect}

\subsection{Theory of Dipole Moments}%

%TCIMACRO{\FRAME{ftbpFU}{8.8326cm}{5.9221cm}{0pt}{\Qcb{Allowed parameter
%regions for $\epsilon=d/\mu<2\times10^{-3}$. The arrows indicate the points
%where new channels start.}}{}{fig2.eps}{\special{ language "Scientific Word";
%type "GRAPHIC";  maintain-aspect-ratio TRUE;  display "USEDEF";
%valid_file "F";  width 8.8326cm;  height 5.9221cm;  depth 0pt;
%original-width 6.915in;  original-height 4.8118in;  cropleft "0";
%croptop "1";  cropright "1.0421";  cropbottom "0";
%filename 'fig2.eps';file-properties "XNPEU";}}}%
%BeginExpansion
\begin{figure}
[ptb]
\begin{center}
\includegraphics[
trim=0.000000in 0.000000in -0.291121in 0.000000in,
height=5.9221cm,
width=8.8326cm
]%
{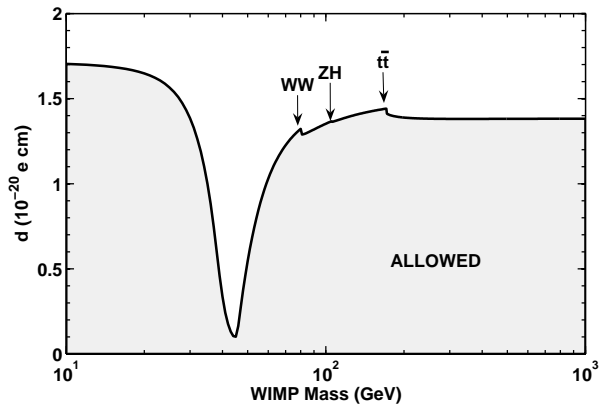}%
\caption{Allowed parameter regions for $\epsilon=d/\mu<2\times10^{-3}$. The
arrows indicate the points where new channels start.}%
\end{center}
\end{figure}
%EndExpansion

The annihilation rates by EDM interaction are suppressed since the s-wave
amplitude is zero due to CP violation, so EDM interaction cannot produce the
right magnitude of the relic density. To reach the thermal average of
annihilation rate $\langle\sigma v_{\text{rel}}\rangle\simeq0.62$ pb, EDM has
to be very large $(\sim10^{-16}e$ $cm)$ and it is ruled out by direct searches
for WIMP near electroweak scale. Simply, we give the relation between EDM and
MDM, $d=\epsilon\mu$, where $\mu$ is MDM, and we consider the minimal Dirac
fermionic dark matter scenario with nonzero MDM \cite{heo09}. With this
relation, the parameter $\epsilon$ becomes a free parameter that would be
constrained by EDM CP related phenomenology. The most simple model
(Lagrangian) is with the complex dipole coupling. If the loop contributions to
all the electromagnetic dipole operators have very similar diagramatic
structure \footnote{Actually this was pointed out in Ref. \cite{Mgr02} for the
ordinary Dirac particles in supersymmetric theories.}, the effective
Lagrangian may be represented with the complex dipole coupling $\mathcal{D}$.%

\begin{equation}
\mathcal{L}_{\mathrm{eff}}=\frac{1}{2}\mathcal{D}\overline{\psi}\sigma_{\mu
\nu}P_{L}\psi F^{\mu\nu}+\frac{1}{2}\mathcal{D}^{\ast}\overline{\psi}%
\sigma_{\mu\nu}P_{R}\psi F^{\mu\nu}=\frac{1}{2}\overline{\psi}\left(
\mu-id\gamma_{5}\right)  \sigma_{\mu\nu}\psi F^{\mu\nu},
\end{equation}
where $P_{L},P_{R}$ are the left and right handed projectors and $F^{\mu\nu}$
is the electromagnetic field strength. The real part of the complex dipole
coupling is consistent with magnetic dipole and the imaginary part is for
electric dipole. The relations with real dipoles are $\mu=\left\vert
\mathcal{D}\right\vert \cos\phi$ and $d$ $=\left\vert \mathcal{D}\right\vert
\sin\phi$, where $\phi$ is the phase of electromagnetic dipole operators. The
relationship between dipoles is given by $d=\mu\tan\phi$ and $\epsilon
=\tan\phi$ would account for all the phenomenology for CP violation of EDM.

The phase of dipole operators would be constrained by the exclusion limit of
EDM \cite{Bcr02} and anomalous MDM for the known Dirac particles like muon or
electron, $\epsilon=$ $\tan\phi\leq2\times10^{-3}$. Our candidate might not be
amenable to this constraint, but we adopt this constraint to analyze EDM
effect for direct searches. Fig.2 shows the allowed parameter regions of EDM
in mass range $10\sim1000$ GeV. The bold line on the border is for
$\epsilon=2\times10^{-3}$, and MDM is taken in Ref.\cite{heo09}, that is
constrained by the relic density.

\subsection{WIMP Detectability}

The cross sections are enhanced by the square of the nuclear charge $Z^{2}$
and this effect might be able to compensate the suppression of EDM CP
violation. Fig.3 shows the regions of the the expected event rates of Eq.(8)
with the estimated EDM from the above agrument. The experimental exclusion
limits of XENON10, CDMS II and SuperCDMS are included. The experimental
sensitivities estimated by the current detectors, XENON10 \cite{xenon08} and
CDMS II \cite{cdm08}, are for 0.1 cpd/kg and the $1\times10^{-4}$ cpd/kg
exclusion limit is considered for the planned detector, SuperCDMS
\cite{scdm06}, respectively. Traditionally, the results of the direct searches
are presented in the form of the WIMP-nucleon cross section at the zero
momentum transfer in the spin independent case, since such normalized forms
are useful to compare results for different types of nuclei. But the cross
section is not defined at the zero momentum transfer in this case, so we
consider the predicted event rates and the experimental exclusion limits for
event rates that estimated for the detectors.%

%TCIMACRO{\FRAME{ftbpFU}{13.25cm}{8.8831cm}{0pt}{\Qcb{The expected event rates
%per kg and day as a function of the WIMP mass. The horizontal lines are the
%exclusion limits.}}{}{fig3.eps}{\special{ language "Scientific Word";
%type "GRAPHIC";  maintain-aspect-ratio TRUE;  display "USEDEF";
%valid_file "F";  width 13.25cm;  height 8.8831cm;  depth 0pt;
%original-width 6.915in;  original-height 4.8118in;  cropleft "0";
%croptop "1";  cropright "1.0406";  cropbottom "0";
%filename 'FIG3.eps';file-properties "XNPEU";}}}%
%BeginExpansion
\begin{figure}
[ptb]
\begin{center}
\includegraphics[
trim=0.000000in 0.000000in -0.280749in 0.000000in,
height=8.8831cm,
width=13.25cm
]%
{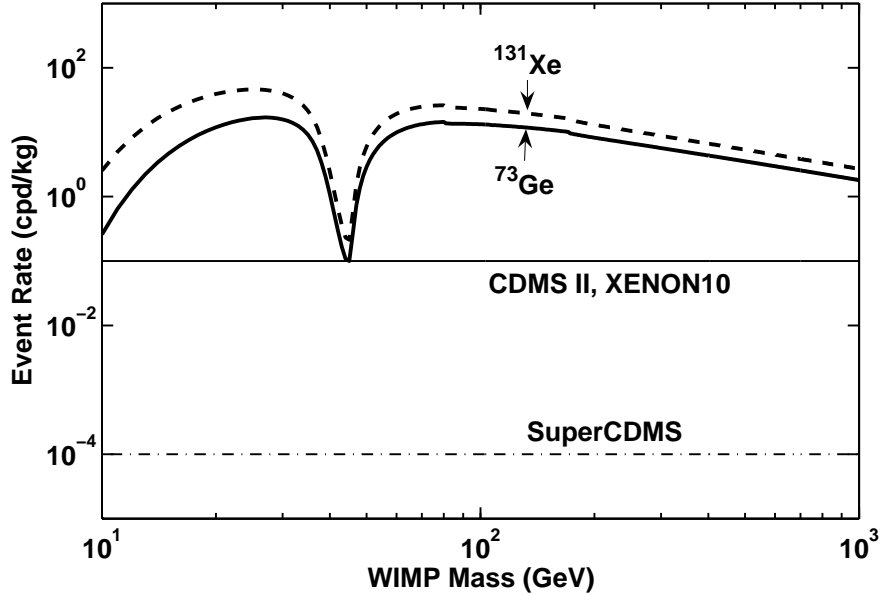}%
\caption{The expected event rates per kg and day as a function of the WIMP
mass. The horizontal lines are the exclusion limits.}%
\end{center}
\end{figure}
%EndExpansion

The predictions are over exclusion limits for $\epsilon=d/\mu=2\times10^{-3}$
and WIMP could be detected in near future by EDM interaction if CP violation
of EDM is not seriously small. The discovery of WIMP by EDM interaction can
also be the key to disclose CP violation nature of EDM.

\section{Conclusion}

The direct limit of electric dipole moment (EDM) and direct search for dark
matter by EDM interaction has been considered in case that the dark matter
candidate is a Dirac particle. The WIMP-nucleus elastic scattering is due to
spin independent interaction, that gives the WIMP electric dipole moment very
strict bound since the WIMP-nucleus elastic scattering cross sections are
enhanced by the square of nuclear charge (number of protons in the nucleus),
$Z^{2}$. The WIMP electric dipole moment constrained by direct searches must
be lower than $7\times10^{-22}e$ $cm$ for WIMP mass of 100 GeV to satisfy the
current experimental exclusion limits. Although we consider that the
interaction of EDM is suppressed by CP violation, the suppression could be
compensated by the enhancement of spin independent interaction. WIMP could
thus be detected by the EDM interaction in near future if the suppression is
not seriously small. The discovery of WIMP by EDM interaction can also be the
key to disclose the nature of CP violation nature.

\end{document}